\documentclass[conference, twocolumn, 10pt]{IEEEtran} 
	\IEEEoverridecommandlockouts
	\usepackage[utf8]{inputenc}
	\usepackage{amsmath}
	\usepackage{amssymb}
	\usepackage{url}
	\usepackage{graphicx}
	\usepackage{epstopdf}
	\usepackage{caption}
	\usepackage{comment}
	\usepackage{subcaption}
	\usepackage{multirow}
	\usepackage{amssymb}
	\usepackage{setspace}
	\usepackage[noadjust]{cite}
	\usepackage{xspace}
	\usepackage{algorithm}
	\usepackage{tabularx}
	\usepackage{changepage}

	\usepackage{graphicx,epsfig,bm}
	\usepackage{cite}
	\usepackage{amsmath,epsfig} 
	\usepackage{times}
	\usepackage{enumerate,type1cm}
	\usepackage{amsfonts,relsize}
	\usepackage{bm}
	\usepackage{amssymb}
	\usepackage{relsize}
	\usepackage{fancybox}
	\usepackage{algorithmic}
	\usepackage{amssymb}
	\usepackage{graphicx,epsfig,bm}
	\usepackage{amsmath,epsfig} 
	\usepackage{times}
	\usepackage{enumerate,type1cm}
	\usepackage{amsfonts,relsize}
	\usepackage{bm}
	\usepackage{amssymb}
	\usepackage{relsize}
	\usepackage{fancybox}
	\usepackage{bbm}
	\usepackage{tikz}
	\usetikzlibrary{fit,positioning}
	\usetikzlibrary{bayesnet}
	\usetikzlibrary{shapes,decorations}
	\usepackage[english]{babel}
	\usepackage{xpatch}
	
	\usepackage{pgf}
	\usepackage{pgfplots}
	\pgfplotsset{compat=1.18}
	\usepackage{tikz}
	\usepackage{subcaption}
	\usepackage{eqparbox}
	\usepackage{makecell}
	\usepackage{svg}
	
	\usepackage[xindy]{glossaries}
	\usepackage{xcolor}
	
	\def\BibTeX{{\rm B\kern-.05em{\sc i\kern-.025em b}\kern-.08em
			T\kern-.1667em\lower.7ex\hbox{E}\kern-.125emX}}

	\newcommand{\norm}[1]{\left\Vert#1\right\Vert}

\newacronym{ai}{AI}{artificial intelligence}
\newacronym{aoa}{AOA}{angle-of-arrival}
\newacronym{aod}{AOD}{angle-of-departure}
\newacronym{ap}{AP}{access point}
\newacronym{asic}{ASIC}{application specific integrated circuit}
\newacronym{awgn}{AWGN}{additive white Gaussian noise}
\newacronym{blue}{BLUE}{best linear unbiased estimator}
\newacronym{ber}{BER}{bit error rate}
\newacronym{bs}{BS}{base station}
\newacronym{ce}{CE}{channel estimation}
\newacronym{cw}{CW}{continuous wave}
\newacronym{cdf}{CDF}{cumulative distribution function}
\newacronym{ced}{CED}{cumulative energy density}
\newacronym{clk}{CLK}{clock}
\newacronym{csi}{CSI}{channel state information}
\newacronym{csp}{CSP}{contact service point}
\newacronym{cpu}{CPU}{central processing unit}
\newacronym{crlb}{CRLB}{Cram\'er-Rao lower bound}
\newacronym{cu}{CU}{central unit}
\newacronym{dc}{DC}{direct current}
\newacronym{dl}{DL}{downlink}
\newacronym{dlt}{DLT}{Distributed Ledger Technology}
\newacronym{dp}{DP}{differencial privacy}
\newacronym{dram}{DRAM}{dynamic random-access memory}
\newacronym{du}{DU}{distributed unit}
\newacronym{ecc}{ECC}{elliptic curve cryptography}
\newacronym{ecdlp}{ECDLP}{elliptic curve discrete logarithm problem}
\newacronym{ecsp}{ECSP}{edge computing service point}
\newacronym{eirp}{EIRP}{equivalent isotropically radiated power}
\newacronym{em}{EM}{electromagnetic}
\newacronym{en}{EN}{energy neutral}
\newacronym{e2e}{E2E}{End-to-End}
\newacronym{fa}{FA}{federation anchor}
\newacronym{fdd}{FDD}{frequency division duplexing}
\newacronym{fhss}{FHSS}{frequency hopping spread spectrum}
\newacronym{fl}{FL}{federated learning}
\newacronym{gamp}{GAMP}{generalized approximate message passing}
\newacronym{gdpr}{GDPR}{general data protection regulation}
\newacronym{gfra}{GFRA}{grant-free random-access}
\newacronym{ia}{IA}{initial access}
\newacronym{ic}{IC}{integrated circuit}
\newacronym{ioe}{IoE}{internet of everything}
\newacronym{iot}{IoT}{internet of things}
\newacronym{id}{ID}{information decoding}
\newacronym{kpi}{KPI}{key performance indicator}
\newacronym{kl}{KL}{Kullback-Leibler}
\newacronym{lca}{LCA}{life cycle assessment}
\newacronym{ls}{LS}{least squares}
\newacronym{lsfc}{LSFC}{large scale fading coefficient}
\newacronym{lis}{LIS}{large intelligent surface}
\newacronym{liion}{Li-ion}{lithium-ion}
\newacronym{llr}{LLR}{log-likelihood ratio}
\newacronym{los}{LoS}{line-of-sight}
\newacronym{lmo}{LMO}{lithium ion manganese oxide}
\newacronym{map}{MAP}{maximum aposteriori}
\newacronym{md}{MD}{miss-detection}
\newacronym{mf}{MF}{matched filter}
\newacronym{ml}{ML}{machine learning}
\newacronym{mle}{MLE}{maximum likelihood estimator}
\newacronym{mmv}{MMV}{multiple-measurement-vector}
\newacronym{mvu}{MVU}{minimum variance unbiased}
\newacronym{mse}{MSE}{mean square error}
\newacronym{mmse}{MMSE}{minimum mean square error}
\newacronym{lmmse}{LMMSE}{linear minimum mean-square error}
\newacronym{ldpc}{LDPC}{low-density parity-check}
\newacronym{mmwave}{mmWave}{millimeter wave}
\newacronym{mimo}{MIMO}{multiple-input multiple-output}
\newacronym{miso}{MISO}{multiple-input single-output}
\newacronym{mpc}{MPC}{multipath component}
\newacronym{mrc}{MRC}{maximum ratio combining}
\newacronym{mrt}{MRT}{maximum ratio transmission}
\newacronym{mtc}{MTC}{machine-type communication}
\newacronym{nmse}{NMSE}{normalized mean square error}
\newacronym{ofdm}{OFDM}{orthogonal frequency-division multiplexing}
\newacronym{ota}{OTA}{over-the-air}
\newacronym{nb}{NB}{narrowband}
\newacronym{nr}{NR}{New Radio}
\newacronym{pa}{PA}{power amplifier}
\newacronym{pdp}{PDP}{power delay profile}
\newacronym{per}{PER}{packet error rate}
\newacronym{pet}{PET}{privacy enhancing technology}
\newacronym{pki}{PKI}{public key infrastucture}
\newacronym{pl}{PL}{path loss}
\newacronym{pc}{PC}{pilot count}
\newacronym{psd}{PSD}{power spectral density}
\newacronym{qrrls}{QR-RLS}{QR decomposition based recursive least squares}
\newacronym{qam}{QAM}{quadrature amplitude modulation}
\newacronym{ra}{RA}{random-access}
\newacronym{rb}{RB}{resource block}
\newacronym{re}{RE}{radio element}
\newacronym{rf}{RF}{radio frequency}
\newacronym{rfid}{RFID}{radio frequency identification}
\newacronym{ris}{RIS}{reflective intelligent surface}
\newacronym{rms}{RMS}{root mean square}
\newacronym{rls}{RLS}{recursive least squares}
\newacronym{rss}{RSS}{received signal strength}
\newacronym{rssi}{RSSI}{received signal strength indicator}
\newacronym{rw}{RW}{RadioWeaves}
\newacronym{sa}{SA}{synchronization anchor}
\newacronym{sbl}{SBL}{sparse-Bayesian-learning}
\newacronym{sdg}{SDG}{Sustainable Development Goal}
\newacronym{sdn}{SDN}{software-defined network}
\newacronym{se}{SE}{spectral efficiency}
\newacronym{ser}{SER}{symbol error rate}
\newacronym{sinr}{SINR}{signal-to-interference-plus-noise ratio}
\newacronym{siso}{SISO}{single-input single-output}
\newacronym{slam}{SLAM}{simultaneous localization and mapping}
\newacronym{snr}{SNR}{signal-to-noise ratio}
\newacronym{srls}{SRLS}{standard recursive least squares}
\newacronym{svd}{SVD}{singular value decomposition}
\newacronym{sp}{SP}{service point}
\newacronym{sram}{SRAM}{static random-access memory}
\newacronym{tdd}{TDD}{time division duplexing}
\newacronym{tdoa}{TDOA}{time-difference-of-arrival}
\newacronym{toa}{TOA}{time-of-arrival}
\newacronym{tx}{TX}{transmitter}
\newacronym{rx}{RX}{receiver}
\newacronym[longplural={user-equipment}]{ue}{UE}{user-equipment}
\newacronym{uhf}{UHF}{ultra high frequency}
\newacronym{ula}{ULA}{uniform linear array}
\newacronym{upa}{UPA}{uniform planar array}
\newacronym{ul}{UL}{uplink}
\newacronym{ura}{URA}{uniform rectangular array}
\newacronym{uwb}{UWB}{ultrawideband}
\newacronym{vb}{VB}{variational-Bayesian}
\newacronym{vbi}{VBI}{variational-Bayesian-inference}
\newacronym{wb}{WB}{wideband}
\newacronym{wpt}{WPT}{wireless power transfer}
\newacronym{zf}{ZF}{zero forcing}

\newacronym{lpt}{LPT}{laser power transfer}
\newacronym{rfpt}{RFPT}{radio frequency power transfer}
\newacronym{ipt}{IPT}{inductive power transfer}
\newacronym{cpt}{CPT}{capacitive power transfer}
\newacronym{apt}{APT}{acoustic power transfer}
\newacronym{cfo}{CFO}{carrier frequency offset}
\newacronym{lo}{LO}{local oscillator}
\newacronym{if}{IF}{intermediate frequency}
\newacronym{duc}{DUC}{digital up conversion}
\newacronym{ddc}{DDC}{digital down conversion}
\newacronym{dsp}{DSP}{digital signal processing}
\newacronym{vco}{VCO}{voltage-controlled oscillator}
\newacronym{pll}{PLL}{phase-locked loop}
\newacronym{lna}{LNA}{low noise amplifier}
\newacronym{dac}{DAC}{digital-to-analog converter}
\newacronym{adc}{ADC}{analog-to-digital converter}
\newacronym{de}{DE}{drain efficiency}
\newacronym{pae}{PAE}{power-added efficiency}
\newacronym{sar}{SAR}{specific absorption rate}
\newacronym{mppt}{MPPT}{maximum power point tracking}
\newacronym{isi}{ISI}{intersymbol interference}
\newacronym{pps}{PPS}{pulse per second}
\newacronym{icnirp}{ICNIRP}{International Commission on Non-Ionizing Radiation Protection}
\newacronym{fd}{FD}{Front-haul distance}
\newacronym{wd}{WD}{Wireless distance}
\newacronym{ntp}{NTP}{Network Time Protocol}
\newacronym{ptp}{PTP}{Precision Time Protocol}
\newacronym{wr}{WR}{White Rabbit}
\newacronym{wlan}{WLAN}{wireless local-area network}

	\long\def\symbolfootnote[#1]#2{\begingroup%
		\def\thefootnote{\fnsymbol{footnote}}\footnote[#1]{#2}\endgroup}
	
	\newcommand{\beq}{\begin{equation}}
		\newcommand{\eeq}{\end{equation}}
	\newcommand{\beqa}{\begin{eqnarray}}
		\newcommand{\eeqa}{\end{eqnarray}}

	\newcommand{\diag}{\mathrm{diag}}

	\usepackage{float}
	\usepackage{colortbl}
	
	\DeclareMathOperator*{\argmin}{arg\,min}
	\DeclareMathOperator*{\argmax}{arg\,max}
	
		\tikzset{
			startstop/.style={
				rectangle, 
				rounded corners,
				minimum width=3cm, 
				minimum height=0.5cm,
				align=center, 
				draw=black, 
			},
			process/.style={
				rectangle, 
				minimum width=3cm, 
				minimum height=0.5cm, 
				align=center, 
				draw=black, 
			},
			decision/.style={
				rectangle, 
				minimum width=3cm, 
				minimum height=0.5cm, align=center, 
				draw=black, 
			},
			arrow/.style={thick,->,>=stealth},
			dec/.style={
				ellipse, 
				align=center, 
				draw=black, 
			},
		}
		
		
		\makeatletter
		\def\BState{\State\hskip-\ALG@thistlm}
		\makeatother
		
		\DeclareCaptionLabelSeparator{periodspace}{.\quad}
			{\end{adjustwidth}}

		\usepackage{stfloats}
			\makeatletter
			\newcommand{\removelatexerror}{\let\@latex@error\@gobble}
			\makeatother
			
	\title{{A Flexible Framework for Grant-Free Random Access in Cell-Free Massive MIMO Systems}
	}
	
	\author{Sai Subramanyam Thoota and Erik G. Larsson\\
		Department of Electrical Engineering (ISY), Link{\"o}ping University, Link{\"o}ping, Sweden\\
		Emails: sai.subramanyam.thoota@liu.se, erik.g.larsson@liu.se\thanks{This work was funded by the REINDEER project of the European Union's Horizon 2020 research and innovation program under grant agreement No.~101013425.}
	 }
	
	\allowdisplaybreaks
	
\begin{document}
	
	\maketitle
	\date{}
	\begin{abstract}
	We propose a novel generalized framework for \gls{gfra} in cell-free massive \acrlong{mimo} systems where multiple geographically separated \glspl{ap} or \glspl{bs} aim to detect sporadically active \glspl{ue}. Unlike a conventional architecture in which all the active \glspl{ue} transmit their signature or pilot sequences of equal length, we admit a flexible pilot length for each \gls{ue}, which also enables a seamless integration into conventional grant-based wireless systems. We formulate the joint \gls{ue} activity detection and the distributed channel estimation as a sparse support and signal recovery problem, and describe a Bayesian learning procedure to solve it. We develop a scheme to fuse the posterior statistics of the latent variables inferred by each \gls{ap} to jointly detect the \glspl{ue}' activities, and utilize them to further refine the channel estimates. In addition, we allude to an interesting point which enables this flexible \gls{gfra} framework to encode the information bits from the active \glspl{ue}. We numerically evaluate the \acrlong{nmse} and the probability of \acrlong{md} performances obtained by the Bayesian algorithm and show that the latent-variable fusion enhances the detection and the channel estimation performances by a large margin. We also benchmark against a genie-aided algorithm which has a prior knowledge of the \glspl{ue}' activities.
	
	\end{abstract}
	
	\begin{keywords}
		\textnormal{Activity detection, cell-free massive \acrshort{mimo}, channel estimation, grant-free random access.
		}
	\end{keywords}
	\section{Introduction}\label{sec:Introduction}
	\Acrfull{gfra} or massive \gls{mtc} is one of the potential technologies that is expected to play a crucial role in the next-generation wireless standards~\cite{Zhilin_TSP_2018,Liang_SPM_2018}. With an increasing demand for several internet-of-things applications in the health, infrastructure, energy sectors, etc., and with a limited availability of wireless resources, obtaining a dedicated block of wireless resources to cater to a sporadically active set of \acrfullpl{ue} may not always be feasible. Moreover, each \gls{mtc} \gls{ue} may not occupy a complete \gls{rb} of the same size as a conventional \gls{ue} which may deem a typical multi-user \acrfull{mimo} receiver algorithm not an appropriate fit. Therefore, it is imperative to develop schemes which are flexible enough to be adapted to both the conventional and \gls{mtc} communications.
	
	Massive \gls{gfra} is a well researched topic where one or multiple \acrfullpl{ap} attempt to detect sporadically active \glspl{ue} using their transmitted \gls{ul} signature or pilot sequences, and a few of the relevant papers are \cite{Zhilin_TSP_2018,Liang_SPM_2018,Kamil_TCOM_2018,Liang_TSP_2018,LiYang_TWC_2019,Unnikrishnan_TCOM_2021,Zhilin_TSP_2021,WeiChen_TWC_2022,LiYang_TWC_2023,HaoZhang_TSP_2024}. Typically, all the existing literature consider a system model where the active \glspl{ue} transmit their pilot sequences of equal length, and the \glspl{ap} detect their activities and estimate their \gls{csi} using sparse signal recovery algorithms~\cite{Tipping_2001_JMLR,Wipf_2004_TSP,Bishop_PRML,Alshoukairi_TSP_2018}. However, to the best of our knowledge, none of the existing papers account for a scenario where the active \glspl{ue} send their signature sequences of different lengths, which can potentially enable a seamless integration with the conventional grant-based wireless systems. This restriction of equal pilot lengths to the \glspl{ue} can possibly narrow down the scope of \gls{gfra} in practical scenarios, especially when there is a resource crunch.
	
	To broaden the realm and make \gls{gfra} appealing to be implemented in practice, we break the notion of equal pilot lengths for all the \gls{mtc} \glspl{ue}, and develop a generalized and a flexible framework in a distributed or cell-free massive \gls{mimo} setup. Therefore, we can fit the \gls{mtc} \glspl{ue} along with any other normal \glspl{ue} which are allocated resources in a grant based manner. The size of the \gls{rb} of a conventional \gls{ue} can be interpreted as the transmission window in which the \gls{mtc} \glspl{ue} can send their pilot sequences of any size. We can see that this also subsumes a standard \gls{gfra} case when the transmission window is of the same size as the pilot length, which makes our structure general.
	
	We propose a novel generalized massive \gls{gfra} framework where the \gls{mtc} \glspl{ue} can transmit their variable length signature sequences within a transmission window of a predefined size for activity detection and channel estimation. We formulate this as a distributed sparse signal and support recovery problem and adopt a Bayesian framework to solve it at the multiple distributed \glspl{ap}. Then, we devise a scheme to fuse the posterior statistics of the latent variables from the \glspl{ap} to improve the activity detection and channel estimation performance. We empirically evaluate the probability of \acrfull{md} and the \acrfull{nmse} performances of the developed algorithm, and benchmark against a genie-aided estimator which has a prior knowledge of the activities of the \glspl{ue}.
	
	We advert to another interesting point that can be utilized by this flexible framework to encode information bits depending on the sizes of the pilot sequence of a \gls{ue} and the transmission window. For instance, if we denote the pilot length and the transmission window by $T$ and $W$, respectively, then we can encode $\lfloor\log_2\left(T-W+1\right)\rfloor$ information bits based on the occupancy of the pilot sequence within the transmission window. Therefore, this generalized structure not only seamlessly integrates into a conventional wireless system but also provides an additional benefit (almost for ``\textit{free}''), which makes it furthermore significant to design specialized and novel receiver algorithms. 
	
	\section{System Model and Problem Statement}\label{sec:SystemModel}
	We consider an \gls{ul} cell-free massive \gls{mimo} wireless communication system with $L$ \glspl{ap} equipped with $N_r$ antennas each, receiving signals from $K$ out of $N$ ($K\ll N$) single transmit antenna \glspl{ue}. We represent the channel between the $k$-th \gls{ue} and the $\ell$-th \gls{ap} by $\mathbf{b}_{\ell k}\in\mathbb{C}^{N_r\times 1}$ which follows a circularly symmetric complex normal distribution $\mathcal{CN}\left(\mathbf{b}_{\ell k};\mathbf{0},\beta_{\ell k}\mathbf{I}_{N_r}\right)$, where $\beta_{\ell k}$ is the \gls{lsfc}. The \glspl{ap} do not have any prior information about the activity pattern of the \glspl{ue}, i.e., they do not know the value of $K$. The $k$-th \gls{ue} can potentially transmit its pilot signal of length $T_{k}$ symbols within an observation window of $W$ symbol intervals. We consider the case when the \glspl{ap} know the size of the transmission window and the lengths of the \glspl{ue}' pilot signals, but do not have any information about the exact starting symbol intervals of the \glspl{ue}' transmissions. We represent the received signal at the $\ell$-th \gls{ap} during the $W$ symbol intervals as:
	\begin{align}
	    \mathbf{Z}_{\ell}^T  
	    &=\sum_{k=1}^K \mathbf{b}_{\ell i_k} \begin{bmatrix}
	        \underbrace{0,\ldots,0}_{S_{i_k}-1} & \mathbf{x}_{i_k}^T & \underbrace{0,\ldots,0}_{W-\left(T_{{i_k}}+S_{i_k}-1\right)}
	    \end{bmatrix}
	    + \mathbf{W}_{\ell}^T,\label{eqn:Y_eqn1}
	\end{align}
	where $\mathbf{Z}_{\ell}^T\in\mathbb{C}^{N_r \times W}$, $\ell\in\{1,\ldots,L\}$, $i_k\in\{1,\ldots,N\}$ denotes the index of the $k$-th active \gls{ue}, $k\in\{1,\ldots,K\}$, ${\mathbf{x}_{i_k}}=\left[x_{{i_k},1},\ldots,x_{{i_k},T_{{i_k}}}\right]^T\in\mathbb{C}^{T_{{i_k}}\times 1}$, and $S_{i_k}\in\{1,\ldots,W-T_{{i_k}}+1\}$ are the pilots and the starting symbol index of the ${i_k}$-th active \gls{ue}, respectively, $\mathbf{W}_{\ell}^T\in\mathbb{C}^{N_r\times W}$ is the additive noise with i.i.d. circularly symmetric complex Gaussian entries of mean $0$ and variance $1$. The maximum transmit power of each \gls{ue} is $P_{\rm max}^{(\rm Tx)}$, i.e., $\mathbb{E}[\norm{\mathbf{x}_k}^2]\le P_{\rm max}^{(\rm Tx)}$, $k\in[1,\ldots,N]$.
	
	Our goal is to develop a distributed algorithm that jointly detects the activity of the \glspl{ue} and estimate their channels given the received signals $\mathbf{Z}_{\ell}$, $\ell\in[1,\ldots,L]$ and the pilot symbols $\mathbf{x}_k$, $k\in\{1,\ldots,N\}$. We rewrite \eqref{eqn:Y_eqn1} to formulate the device activity detection and channel estimation problem as follows:
	\begin{align}
	    \mathbf{Z}_{\ell}^T &= \sum_{k=1}^K \begin{bmatrix}
	        \mathbf{0}_{N_r\times \left(S_{i_k}-1\right)}&\mathbf{b}_{\ell i_k}&\mathbf{0}_{N_r\times \left(W-T_{i_k}-S_{i_k}+1\right)}
	    \end{bmatrix} \nonumber\\&\qquad\qquad\qquad \times \begin{bmatrix}
	        \mathbf{x}_{i_k}^T & 0&\ldots&0\\
	        \vdots&\vdots&\ddots&\vdots\\
	        0&0&\ldots& \mathbf{x}_{i_k}^T 
	    \end{bmatrix}+ \mathbf{W}_{\ell}^T,\label{eqn:Y_eqn21}\\
	    &=\sum_{k=1}^K \mathbf{G}_{\ell i_k}^T \mathbf{D}_{i_k}^T+\mathbf{W}_{\ell}^T,\label{eqn:Y_eqn22}
	\end{align}
	where $\mathbf{G}_{\ell i_k}\in\mathbb{C}^{\left(W-T_{i_k}+1\right)\times N_r}$ contains one non-zero row ($S_{i_k}$-th row is non-zero), and $\mathbf{D}_{i_k}^T\in\mathbb{C}^{\left(W-T_{i_k}+1\right)\times W}$ contains the pilot sequences of the $i_k$-th \gls{ue}. Now, when we include all the inactive devices into \eqref{eqn:Y_eqn22}, we get
	\begin{align}
	    \mathbf{Z}_{\ell}^T &= \sum_{k=1}^N \mathbf{G}_{\ell k}^T \mathbf{D}_{k}^T+\mathbf{W}_{\ell}^T,\label{eqn:Y_eqn3}
	\end{align}
	where $\mathbf{G}_{\ell k}\in\mathbb{C}^{\left(W-T_{k}+1\right)\times N_r}$ is non-zero only for the $K$ active \glspl{ue} (which is unknown). We transpose \eqref{eqn:Y_eqn3} and rewrite in a matrix form as:
	\begin{align}
	    \mathbf{Z}_{\ell} = \mathbf{D} \mathbf{G}_{\ell}+\mathbf{W}_{\ell},\qquad\ell\in[1,\ldots,L],\label{eqn:Y_eqn4}
	\end{align}
	where \\$\mathbf{G}_{\ell}=\begin{bmatrix}
	    \mathbf{G}_{\ell 1}^T&\ldots&\mathbf{G}_{\ell N}^T
	\end{bmatrix}^T\in\mathbb{C}^{\left((W+1)N-\sum_{k=1}^N T_{k}\right)\times N_r}$ is a joint row-sparse matrix with $K$ non-zero rows, and \\$\mathbf{D} = \begin{bmatrix}
	    \mathbf{D}_1&\ldots&\mathbf{D}_N
	\end{bmatrix}\in\mathbb{C}^{W\times\left((W+1)N-\sum_{k=1}^N T_{k}\right)}$ is the over-complete dictionary matrix. 
	
	Estimating $\mathbf{G}_{\ell}$ using the received signal $\mathbf{Z}_{\ell}$ is a local \gls{mmv} sparse signal recovery problem at the $\ell$-th \gls{ap}. 
	Our aim is to use the received signals $\mathbf{Z}_1,\ldots,\mathbf{Z}_L$ to jointly detect the active \glspl{ue} and estimate the channels at each \gls{ap}. Moreover, the support set of $\{\mathbf{G}_1,\ldots,\mathbf{G}_L\}$ also determines the starting symbol times of the active \glspl{ue}. For convenience, we denote the $n$-th column of $\mathbf{G}_{\ell}$ and $\mathbf{Z}_{\ell}$ as $\mathbf{g}_{\ell n}\in\mathbb{C}^{\left((W+1)N-\sum_{k=1}^N T_{k}\right)\times 1}$ and $\mathbf{z}_{\ell n}\in\mathbb{C}^{W\times 1}$, $n\in\{1,\ldots,N_r\}$, respectively, and $M\triangleq(W+1)N-\sum_{k=1}^N T_{k}$. Now, we describe the algorithm for the \gls{ue} activity detection and distributed channel estimation.

	\section{Sparse Bayesian Joint User Activity Detection and Distributed Channel Estimation}
	Our goal is to infer the posterior distributions of $\mathbf{G}_{\ell}$ given $\mathbf{Z}_{\ell}$ and $\mathbf{x}_k$, $k\in\{1,\ldots,N\}$, $\ell\in\{1,\ldots,L\}$. Each \gls{ap} uses its local received signal to estimate its channel in a distributed fashion and a \gls{cpu} fuses the posterior statistics of the latent variables to jointly detect the active \glspl{ue}. A key point to note is that the support set of $\mathbf{G}_{\ell}$, $\ell\in\{1,\ldots,L\}$, is common across all the \glspl{ap}, and we utilize this for the joint \gls{ue} activity detection. We adopt the \gls{sbl} or \gls{vb} framework to infer the posterior distributions of the \glspl{ue}' channels locally at each \gls{ap}, and then combine them to detect their activities at a \gls{cpu}. We refer the reader to \cite{Tipping_2001_JMLR,Wipf_2004_TSP,Bishop_PRML} for detailed descriptions of \gls{vb} and \gls{sbl}.
	
	We describe the \gls{vb} procedure here. To begin with, we impose a sparsity promoting two-stage hierarchical complex Gaussian prior on the columns of $\mathbf{G}_{\ell}$ with mean $\mathbf{0}$ and a common diagonal precision matrix $\mathbf{P}_{\ell}=\diag\left(\boldsymbol{\alpha}_{\ell}\right)$, where $\boldsymbol{\alpha}_{\ell} = \left[\alpha_1,\ldots,\alpha_M\right]^T\in\mathbb{R}_+^{M\times 1}$. We treat the elements of $\boldsymbol{\alpha}_{\ell}$ as independent and identically distributed latent variables which follow a non-informative Gamma hyperprior with given rate and shape parameters. Such a two-stage hierarchical structure results in a Student's $t$-distributed prior on $\mathbf{g}_{\ell n}$, $n\in\{1,\ldots,N_r\}$, $l\in\{1,\ldots, L\}$, which is heavy tailed and promotes sparse estimates.
	
	\subsection{Distributed Channel Estimation}\label{sec:VBChEst}
	To obtain the posterior distributions of the channels via the \gls{vb} procedure, we impose an independence structure on the latent variables $\mathbf{g}_{\ell n}$, $n\in\{1,\ldots,N_r\}$, and $\alpha_{\ell m}$, $m\in\{1,\ldots,M\}$, $\ell\in\{1,\ldots,L\}$, which means that the joint posterior probability distribution is the product of the posterior distributions of the individual latent variables. Then, we compute them to minimize the \gls{kl} divergence between the approximate and the original posterior distributions. We do not include the fundamentals of the \gls{vb} or \gls{sbl} framework due to lack of space. In \gls{vb}, we decompose the natural logarithm of the joint probability distribution of the observations and the latent variables as follows:
	\begin{align}
	    &\log P\left(\mathbf{Z}_{\ell},\mathbf{G}_{\ell},\boldsymbol{\alpha}_{\ell};{\mathbf{D}},a,b\right)
	    \nonumber\\
	        &=\sum_{n=1}^{N_r} \log P_{\mathbf{z}_{\ell n}}\left(\mathbf{z}_{\ell n}\,\arrowvert\,\mathbf{g}_{\ell n};{\mathbf{D}}\right)
	        + \sum_{n=1}^{N_r} \log P_{\mathbf{g}_{\ell n}}\left(\mathbf{g}_{\ell n}\,\arrowvert\,\boldsymbol{\alpha}_{\ell}\right)\nonumber\\   &\qquad\qquad\qquad\qquad+ \sum_{m=1}^{M} \log P_{\alpha_{\ell m}}\left(\alpha_{\ell m};a,b\right),\label{eqn:JointProbDist1}
	    \end{align}
	where the conditional probability distributions are given by
	\begin{align}
	    &P_{\mathbf{z}_{\ell n}}\left(\mathbf{z}_{\ell n}\,\arrowvert\,\mathbf{g}_{\ell n};{\mathbf{D}}\right) \propto  \exp\left(-\norm{\mathbf{z}_{\ell n}-\mathbf{D}\mathbf{g}_{\ell n}}^2\right),\label{eqn:CondProbzln}\\
	   &P_{\mathbf{g}_{\ell n}}\left(\mathbf{g}_{\ell n}\,\arrowvert\,\boldsymbol{\alpha}_{\ell}\right)\propto \exp\left(-\mathbf{g}_{\ell n}^H \mathbf{P}_{\ell}\mathbf{g}_{\ell n}+\sum_{m=1}^M \log\,\alpha_{\ell m}\right), \label{eqn:CondProbgln}\\
	   &P_{\alpha_{\ell m}}\left(\alpha_{\ell m};a,b\right) \propto  \alpha_{\ell m}^{a-1} \exp\left(-b\,\alpha_{\ell m}\right),\label{eqn:CondProbalphal}
	\end{align}
	where $a$ and $b$ are the shape and rate parameters of the hyperprior $P_{\alpha_{\ell m}}\left(\alpha_{\ell m};a,b\right)$, respectively. We exclude the constant terms in \eqref{eqn:CondProbzln}, \eqref{eqn:CondProbgln} and \eqref{eqn:CondProbalphal} for brevity. Now, we compute the posterior probability distribution ${q}_{\mathbf{g}_{\ell n}}(\mathbf{g}_{\ell n})$ given the observations as:
	\begin{align}
	    \log\,{q}_{\mathbf{g}_{\ell n}}\left(\mathbf{g}_{\ell n}\right)\propto \left<\log P\left(\mathbf{Z}_{\ell},\mathbf{G}_{\ell},\boldsymbol{\alpha}_{\ell};{\mathbf{D}},a,b\right)\right>_{\sim{q}_{\mathbf{g}_{\ell n}}},\label{eqn:logqgln}
	\end{align}
	where the operator $\left<\cdot\right>_{\sim{q}_{\mathbf{g}_{\ell n}}}$ denotes the expectation with respect to the posterior distributions of all the latent variables except ${q}_{\mathbf{g}_{\ell n}}$. Substituting \eqref{eqn:CondProbzln}, \eqref{eqn:CondProbgln}, and \eqref{eqn:CondProbalphal} in \eqref{eqn:JointProbDist1}, and grouping the terms dependent on $\mathbf{g}_{\ell n}$ together, we get
	\begin{align}
	    &\log\,{q}_{\mathbf{g}_{\ell n}}\left(\mathbf{g}_{\ell n}\right)\nonumber\\ 
	        &\propto-\left(\mathbf{g}_{\ell n}^H\left({\mathbf{D}}^H{\mathbf{D}}+\left<\mathbf{P}_{\ell}\right>\right)\mathbf{g}_{\ell n}-2\Re\left(\mathbf{g}_{\ell n}^H{\mathbf{D}}^H\mathbf{z}_{\ell n}\right)\right).\label{eqn:qgln1}
	\end{align}
	Using the structure of \eqref{eqn:qgln1}, we deduce that ${q}_{\mathbf{g}_{\ell n}}$ follows a complex normal distribution with the covariance and mean:
	\begin{align}
	    \boldsymbol{\Sigma}_{\ell} &=\left({\mathbf{D}}^H{\mathbf{D}}+\left<\mathbf{P}_{\ell}\right>\right)^{-1},\label{eqn:qglnCov}\\
	    \left<\mathbf{g}_{\ell n}\right>&=\boldsymbol{\Sigma}_{\ell} {\mathbf{D}}^H\mathbf{z}_{\ell n},\qquad n\in\{1,\ldots,N_r\},\label{eqn:qglnMean}
	\end{align}
	respectively. To compute $\boldsymbol{\Sigma}_{\ell}$ with a reduced complexity, we use the Woodbury matrix inversion lemma to get
	\begin{align}
	    \boldsymbol{\Sigma}_{\ell} &=\left<\mathbf{P}_{\ell}\right>^{-1}\nonumber\\&\quad\times\left[\mathbf{I}_M-\mathbf{D}^H\left(\mathbf{I}_W+\mathbf{D}\left<\mathbf{P}_{\ell}\right>^{-1}\mathbf{D}^H\right)^{-1}\mathbf{D}\left<\mathbf{P}_{\ell}\right>^{-1}\right]\!\!,\label{eqn:qglnCov_ReducedComplexity}
	\end{align}
	where $\left<\mathbf{P}_{\ell}\right>$ is the posterior mean of $\mathbf{P}_{\ell}$, $\mathbf{I}_W$ and $\mathbf{I}_M$ are identity matrices of size $W\times W$ and $M\times M$, respectively. From \eqref{eqn:qglnMean}, we see that the covariance matrix $\boldsymbol{\Sigma}_{\ell}$ is common for the posterior distributions of the channels across all the $N_r$ antennas, and we write the posterior mean of $\mathbf{G}_{\ell}$ as:
	\begin{align}
	    \left<\mathbf{G}_{\ell}\right>&=\boldsymbol{\Sigma}_{\ell} {\mathbf{D}}^H\mathbf{Z}_{\ell}, \qquad \ell\in\{1,\ldots,L\}.\label{eqn:qglnMeanMat}
	\end{align}
	Now we shift to the derivation of the posterior distribution of $\boldsymbol{\alpha}_{\ell}$ which is used to detect the \glspl{ue}' activities.
	\subsection{Latent-variable-Fusion based User Activity Detection} 
	We derive the posterior distributions ${q}_{\alpha_{\ell m}}(\alpha_{\ell m})$ of $\alpha_{\ell m}$, $m\in\{1,\ldots,M\}$, $\ell\in\{1,\ldots,L\}$, and then propose a mechanism to fuse their statistics to detect the \glspl{ue}' activities. We compute the expectation of the natural logarithm of the joint probability distribution in \eqref{eqn:JointProbDist1} with respect to the posterior distributions of all the latent variables except $\alpha_{\ell m}$ to obtain ${q}_{\alpha_{\ell m}}(\alpha_{\ell m})$ as:
	\begin{align}
	    \log\,{q}_{\alpha_{\ell m}}\left(\alpha_{\ell m}\right)\propto \left<\log P\left(\mathbf{Z}_{\ell},\mathbf{G}_{\ell},\boldsymbol{\alpha}_{\ell};{\mathbf{D}},a,b\right)\right>_{\sim{q}_{\alpha_{\ell m}}},\nonumber
	\end{align}
	where the operator $\left<\cdot\right>_{\sim{q}_{\alpha_{\ell m}}}$ denotes the expectation with respect to the posterior probability distributions of all the latent variables except ${q}_{\alpha_{\ell m}}$. Grouping the terms dependent on $\alpha_{\ell m}$ together and simplifying, we get
	\begin{align}
	    &\log\,{q}_{\alpha_{\ell m}}\left(\alpha_{\ell m}\right)\propto \left(a+N_r-1\right)\log \alpha_{\ell m}\nonumber\\&\qquad\qquad\qquad\qquad\quad-\left(b+\sum_{n=1}^{N_r} \left<\left\arrowvert g_{\ell n m}\right\arrowvert^2\right>\right)\alpha_{\ell m},\label{eqn:logqalphalm}
	\end{align}
	where $g_{\ell n m}$ is the $m$-th element of $\mathbf{g}_{\ell n}$, $m\in\{1,\ldots,M\}$, and $\left<\left\arrowvert g_{\ell n m}\right\arrowvert^2\right>$ can be computed using $\left<\mathbf{G}_{\ell}\right>$ and $\boldsymbol{\Sigma}_{\ell}$. From the structure of \eqref{eqn:logqalphalm}, we deduce that ${q}_{\alpha_{\ell m}}\left(\alpha_{\ell m}\right)$ follows a Gamma distribution with its posterior mean $\left<\alpha_{\ell m}\right>$ as follows:
	\begin{align}
	        \left<\alpha_{\ell m}\right>=\frac{a+N_r}{b+\sum_{n=1}^{N_r} \left<\left\arrowvert g_{\ell nm}\right\arrowvert^2\right>}, \quad m\in\{1,\ldots,M\}.\label{eqn:qalphalm_mean}
	\end{align}
	As we mentioned before, the support sets of $\mathbf{G}_{\ell}$, $\ell\in\{1,\ldots,L\}$, are common across all the \glspl{ap}. Therefore, we utilize the posterior means obtained in \eqref{eqn:qalphalm_mean} at each \gls{ap} to jointly detect the \glspl{ue}' activities. 
	
	We apply a weighted-average method to fuse the posterior means $\left<\alpha_{\ell m}\right>$, $\ell\in\{1,\ldots,L\}$ and $m\in\{1,\ldots,M\}$ sent from the \glspl{ap} to the \gls{cpu}. We assume that the \gls{cpu} has an estimate of $\beta_{\ell k}$, $k\in\{1,\ldots,N\}$, $\ell\in\{1,\ldots,L\}$ using the history of the \glspl{ue}' activities. For each \gls{ue}, the \gls{cpu} selects the \gls{ap} with the largest \gls{lsfc} as its master \gls{ap}, i.e.,
	\begin{align}
	    \ell_k^{(\rm M)} = \argmax_{l\in\{1,\ldots,L\}} \beta_{\ell k},\qquad k\in\{1,\ldots,N\},\label{eqn:MasterAPSelect}
	\end{align}
	\setlength{\textfloatsep}{0pt}
	\begin{algorithm}[!t]
		\caption{User Activity Detection and Channel Estimation} \label{Alg:UADCE}
		\begin{algorithmic}[1]
			\renewcommand{\algorithmicrequire}{\textbf{Input:}}
			\renewcommand{\algorithmicensure}{\textbf{Output:}}
			\REQUIRE $\mathbf{Z}_{\ell}$, $\beta_{\ell k}$, $k\in\{1,\ldots,N\}$, $\ell\in\{1,\ldots,L\}$, $\mathbf{D}$, $a$, $b$.
			\ENSURE  $\left<\mathbf{G}_1\right>,\ldots,\left<\mathbf{G}_L\right>$, $\left<\bar{\boldsymbol{\alpha}}\right>$.
			\STATE \textbf{Initialize} 
	            $\left<\bar{\boldsymbol{\alpha}}_{\ell}\right>$, $\ell\in\{1,\ldots,L\}$.
	            \STATE Select master \glspl{ap} $\ell_k^{(\rm M)}$, $k\in\{1,\ldots,N\}$ using \eqref{eqn:MasterAPSelect}.
			\REPEAT
	            \STATE $\left<\boldsymbol{\alpha}_{\ell}\right> = \left<\bar{\boldsymbol{\alpha}}_{\ell}\right>$.
	            \REPEAT
	            \STATE $\left<\mathbf{P}_{\ell}\right> = \diag\left(\left<\boldsymbol{\alpha}_{\ell}\right>\right)$, $\ell\in\{1,\ldots,L\}$.
			\STATE Compute $\boldsymbol{\Sigma}_{\ell}$ and $\left<\mathbf{G}_{\ell}\right>$, $\ell\in\{1,\ldots,L\}$ using \eqref{eqn:qglnCov_ReducedComplexity} and \eqref{eqn:qglnMeanMat}, respectively.
	            \STATE Compute $\left<\boldsymbol{\alpha}_{\ell}\right>$, $\ell\in\{1,\ldots,L\}$ using \eqref{eqn:qalphalm_mean}.
	            \UNTIL {\rm MAX\_ITER}
	            \FOR {$k=1$ to $N$}
	            \FOR {$m=\left(W+1\right)\left(k-1\right)-\sum_{k'=1}^{k-1}T_{k'}+1$ to $\left(W+1\right)k-\sum_{k'=1}^{k}T_{k'}$}
	            \STATE Compute $\left<\bar{\alpha}_{\ell m}\right>$, $\ell\in\{1,\ldots,L\}$ using \eqref{eqn:alphabarlm_withCSI}.
	            \STATE $\left<\bar{\alpha}_m\right> = \left<\bar{\alpha}_{\ell_k^{(\rm M)}m}\right>$.
	            \ENDFOR
	            \ENDFOR
			\UNTIL stopping condition is met	
		\end{algorithmic}
	\end{algorithm}
	We depend on the fact that a \gls{ue}'s activity statistic computed at its master \gls{ap} is more reliable than that obtained at the other \glspl{ap}. Therefore, we allot a higher weight to the posterior means determined at the master \glspl{ap} than that of the other \glspl{ap} for each \gls{ue}. Once the \glspl{ap} compute $\left<\boldsymbol{\alpha}_{\ell}\right>$, $\ell\in\{1,\ldots,L\}$, they feed them back to the \gls{cpu} which computes the following:
	\begin{align}
	    &{\left<\bar{\alpha}_{\ell m}\right>} = \frac{\beta_{\ell k}\left<{\alpha}_{\ell m}\right>+\beta_{\ell_k^{(\rm M)} k}\left<\alpha_{\ell_k^{(\rm M)} m}\right>}{\beta_{\ell k}+\beta_{\ell_k^{(\rm M)} k}},\quad k\in\{1,\ldots,N\},\label{eqn:alphabarlm_withCSI}\\
	    & m\in\left\{\left(W+1\right)\left(k-1\right)-\sum_{k'=1}^{k-1}T_{k'}+1,\ldots,\right.\nonumber\\&\qquad\qquad\qquad\qquad\qquad\left.\left(W+1\right)k-\sum_{k'=1}^{k}T_{k'}\right\}.\label{eqn:alphabarlm_withCSI_mrange}
	\end{align}
	Note that the indices $m$ in \eqref{eqn:alphabarlm_withCSI_mrange} vary for different \glspl{ue} depending on the lengths of their signature sequences. The \gls{cpu} then forwards  $\left<\bar{\boldsymbol{\alpha}}_{\ell}\right>\triangleq\left[\left<\bar{\alpha}_{\ell 1}\right>\ldots\left<\bar{\alpha}_{\ell M}\right>\right]$ to the $\ell$-th \gls{ap}, $\ell\in\{1,\ldots,L\}$. We mention that if $\ell=\ell_k^{(M)}$ for any $k$, then $\left<\bar{\alpha}_{\ell m}\right> = \left<\alpha_{\ell_k^{(\rm M)} m}\right>$ for all $m$ in the range given in \eqref{eqn:alphabarlm_withCSI_mrange}.
	
	From \eqref{eqn:qglnCov_ReducedComplexity}, \eqref{eqn:qglnMeanMat} and \eqref{eqn:qalphalm_mean}, we see that the posterior statistics of the latent variables are interdependent on each other. Therefore, we initialize them randomly and execute the \gls{vb} procedure iteratively till it converges to a locally optimal solution. We present the method for the joint \gls{ue} activity detection and the distributed channel estimation in the Algorithm~\ref{Alg:UADCE}. To obtain reliable estimates of $\left<\bar{\boldsymbol{\alpha}}_{\ell}\right>$, $\ell\in\{1,\ldots,L\}$, we execute the steps $6$ to $8$ for MAX\_ITER iterations (set to $5$ in our simulations) and then fuse the posterior statistics to detect the \glspl{ue}. Once we obtain $\left<\bar{\boldsymbol{\alpha}}_{\ell}\right>$, $\ell\in\{1,\ldots,L\}$ at the \gls{cpu} upon convergence, we reshape it based on the indices in \eqref{eqn:alphabarlm_withCSI_mrange} for all the \glspl{ue}. Then, for each \gls{ue}, we choose 
	\begin{align}
	    m_k^{(\rm min)} = \argmin_{m\,\in\,{\textrm{Eqn. }\eqref{eqn:alphabarlm_withCSI_mrange}}} {\left<\bar{\alpha}_{\ell_k^{(\rm M)} m}\right>},\quad k\in\{1,\ldots,N\},\label{eqn:UserDet}
	\end{align}
	to obtain $\boldsymbol{\alpha}^{(\rm min)}\triangleq\left[{\left<\bar{\alpha}_{\ell_1^{(\rm M)} m_1^{(\rm min)}}\right>},\ldots,{\left<\bar{\alpha}_{\ell_N^{(\rm M)} m_N^{(\rm min)}}\right>}\right]^T\in\mathbb{R}_+^{N\times 1}$. Then, we set a threshold value (set to $1$ in our simulations), and select all the indices of the entries in $\boldsymbol{\alpha}^{(\rm min)}$ which are below it as the detected \glspl{ue}. Note that the entries corresponding to the active \glspl{ue} in \eqref{eqn:UserDet} capture their starting symbol intervals.
	
	To reduce the complexity of computing the matrix inverse in \eqref{eqn:qglnCov_ReducedComplexity}, we also include a \gls{gamp} based variant of \gls{vb} in our simulations~\cite{Alshoukairi_TSP_2018}. We omit the complete mathematical description, convergence, and computational complexity analysis due to lack of space. 
	
	\begin{figure}[!]
		\centering
		\includegraphics[width=\linewidth]{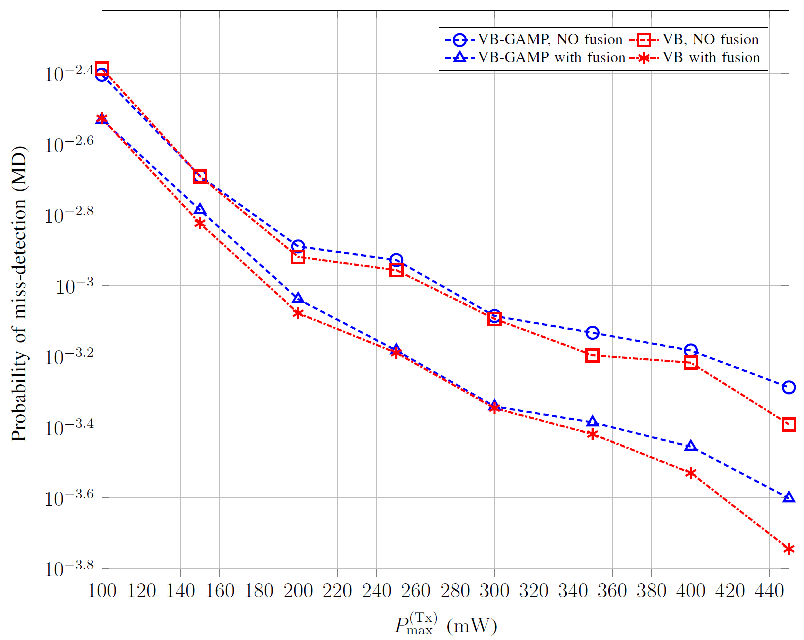}
		\caption{Probability of \gls{md} as a function of $P_{\rm max}^{(\rm Tx)}$.}
		\label{fig:ProbMD}
		\vspace{-0.05cm}
	\end{figure}
	
	\begin{figure}[!]
		\centering
		\includegraphics[width=\linewidth]{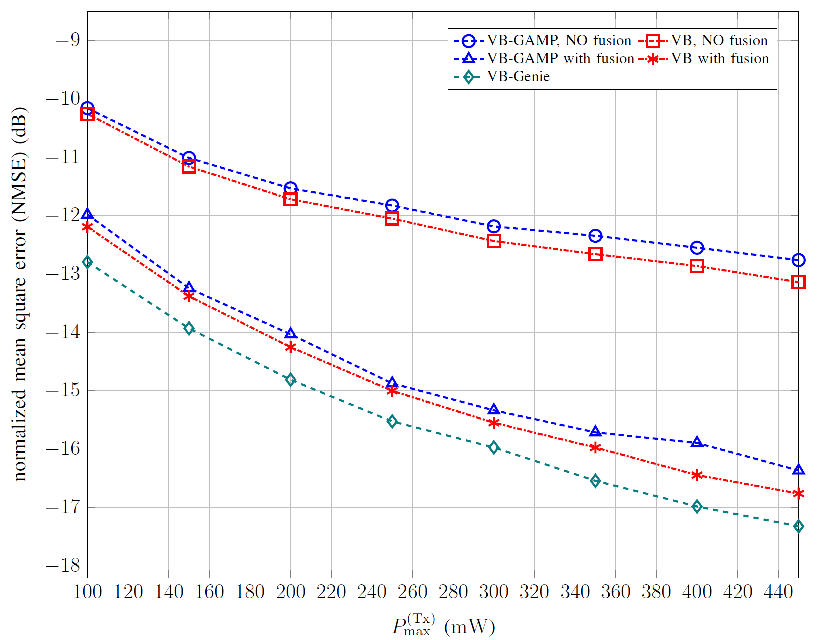}
		\caption{\gls{nmse} as a function of $P_{\rm max}^{(\rm Tx)}$.}
		\label{fig:NMSE}
	\end{figure}

	\section{Simulation Results}\label{sec:SimResults}
	We numerically evaluate the probability of \acrfull{md} and the \acrfull{nmse} performances to demonstrate the efficacy of the proposed flexible \gls{gfra} framework. We deploy $20$ \glspl{ap} with $2$ antennas each at a height of $10$~m and $200$ \glspl{ue} uniformly at random in a square area of $1\times 1$~${\rm km}^2$. We set the number of active \glspl{ue} to $20$. We use a wrap-around technique to approximately generate an infinitely large network with $40$ antennas and $20$ active \glspl{ue} per ${\rm km}^2$~\cite{Marzetta_Fundamentals_2016,Ozlem_Cellfree_2021}. We sample the pilot lengths from a uniform distribution between $20$ and $24$ ($W=24$). We use complex Gaussian pilots of unit energy and a bandwidth of $1$~MHz. The noise power spectral density is set to $-169$~{\rm dBm}, and the \gls{lsfc} (in {\rm dB}) of the channel between the $k$-th \gls{ue} and the $\ell$-th \gls{ap} is generated as: $\beta_{\ell k}=-140.6-36.7\log_{10}\left(d_{\ell k}\right)+\Psi_{\ell k}$, where $d_{\ell k}$ is the distance in {\rm km}, $\Psi_{\ell k}$ is the log-normal shadowing distributed as $\mathcal{N}(0,\sigma_{\rm sf}^2)$ and $\sigma_{\rm sf}$ is set to $4$~{\rm dB}~\cite{LTE2010b}. We set $a$ and $b$ to $10^{-10}$.
	
	We vary $P_{\rm max}^{\rm (Tx)}$ from $100$~{\rm mW} to $450$~{\rm mW} in our simulations. 
	We adopt the \gls{ue}-centric power allocation scheme developed in \cite{Jianan_WCOML_2022} to set the instantaneous transmit power (denoted $P_{k}^{\rm (Tx)}$) of the $k$-th \gls{ue} as: 
	$    P_{k}^{\rm (Tx)} = P_{\rm max}^{\rm (Tx)}\frac{\min_{k'}\beta_{\ell_{k'}^{(\rm M)}k'}}{\beta_{\ell_{k}^{(\rm M)}k}}$, $k\in\{1,\ldots,K\}$. 
	This scheme makes sure only the \glspl{ue} with the least \gls{lsfc} to any \gls{ap} transmits with the maximum power ensuring fairness among them. We set the maximum number of iterations for all the algorithms to $250$, which means that the Steps $4$ to $15$ in the Algorithm~\ref{Alg:UADCE} are executed either till convergence or for a maximum of $\left\lfloor\frac{250}{\rm MAX\_ITER}\right\rfloor$ iterations.
	
	Fig.~\ref{fig:ProbMD} compares the probability of \gls{md} of the distributed latent-variable fusion based \gls{ue} activity detection and the state-of-the-art distributed \gls{vb} based procedures (with legend entries ``NO fusion": Steps $4$ and $10$ to $15$ in Algorithm~\ref{Alg:UADCE} are not executed for this). This figure shows that our proposed flexible framework with variable pilot lengths is able to achieve a good \gls{ue} detection performance irrespective of whether we use a latent variable fusion or not. Moreover, we see that the detection probability improves by a large margin by combining the posterior statistics of the latent variables at the \gls{cpu} (legend entries ``with fusion''). For example, the maximum transmit power of the \glspl{ue} can be reduced by around $100$~mW to achieve a probability of \gls{md} of $0.08~\%$ using the latent variable fusion approach, which leads to large power savings to the \glspl{ue}. This also shows that the posterior statistics obtained at the master \glspl{ap} are more reliable than that of the other \glspl{ap}, which improves the detection rate.
	
	Fig.~\ref{fig:NMSE} depicts the \gls{nmse} performances of the distributed \gls{vb} channel estimation procedures with and without latent variable fusion. We also compare against a genie-aided channel estimator which has a prior knowledge about the active \glspl{ue}, and therefore can be considered as a lower bound for the \gls{nmse} of the developed latent variable fusion based estimators. We clearly see that by utilizing the posterior statistics sent from the \glspl{ap} to the \gls{cpu}, the performance moves close to the lower bound. For instance, we can achieve the same \gls{nmse} of VB-Genie for the developed distributed \gls{vb} algorithm by expending only a little less than $50$~mW of the maximum transmit power. Moreover, from the Figures~\ref{fig:ProbMD} and \ref{fig:NMSE}, we observe that a \gls{vb}-\gls{gamp} implementation results in a marginal loss in performance compared to \gls{vb} procedure.
	
	\section{Conclusions \& Future Work}\label{sec:Conclusions}
	We proposed a novel flexible framework for \gls{gfra} in cell-free massive \gls{mimo} systems which admitted variable length signature sequences for the \glspl{ue}. We formulated the joint \gls{ue} activity detection and the distributed channel estimation as a sparse support and signal recovery problem, and described a Bayesian learning procedure to solve it. To enhance the detection performance, we devised a latent-variable fusion mechanism to combine the posterior statistics inferred at the \glspl{ap}. We also referred to an intricate point to encode the information bits from the active \glspl{ue} without any additional transmit power. Finally, we numerically evaluated the \gls{nmse} and the \gls{md} performances of the \gls{sbl} algorithm to illustrate the efficacy of the generalized framework. We aim to investigate the joint activity detection, channel estimation, and data detection for the flexible framework as part of our future work.
	
	\bibliographystyle{IEEEtran}
	\bibliography{IEEEabrv,reff}

\begin{thebibliography}{10}
\providecommand{\url}[1]{#1}
\csname url@samestyle\endcsname
\providecommand{\newblock}{\relax}
\providecommand{\bibinfo}[2]{#2}
\providecommand{\BIBentrySTDinterwordspacing}{\spaceskip=0pt\relax}
\providecommand{\BIBentryALTinterwordstretchfactor}{4}
\providecommand{\BIBentryALTinterwordspacing}{\spaceskip=\fontdimen2\font plus
\BIBentryALTinterwordstretchfactor\fontdimen3\font minus
  \fontdimen4\font\relax}
\providecommand{\BIBforeignlanguage}[2]{{%
\expandafter\ifx\csname l@#1\endcsname\relax
\typeout{** WARNING: IEEEtran.bst: No hyphenation pattern has been}%
\typeout{** loaded for the language `#1'. Using the pattern for}%
\typeout{** the default language instead.}%
\else
\language=\csname l@#1\endcsname
\fi
#2}}
\providecommand{\BIBdecl}{\relax}
\BIBdecl

\bibitem{Zhilin_TSP_2018}
Z.~Chen, F.~Sohrabi, and W.~Yu, ``Sparse activity detection for massive
  connectivity,'' \emph{{IEEE} Trans. Signal Process.}, vol.~66, no.~7, pp.
  1890--1904, 2018.

\bibitem{Liang_SPM_2018}
L.~Liu, E.~G. Larsson, W.~Yu, P.~Popovski, C.~Stefanovic, and E.~de~Carvalho,
  ``Sparse signal processing for grant-free massive connectivity: A future
  paradigm for random access protocols in the internet of things,''
  \emph{{IEEE} Signal Process. Mag.}, vol.~35, no.~5, pp. 88--99, 2018.

\bibitem{Kamil_TCOM_2018}
K.~Senel and E.~G. Larsson, ``Grant-free massive mtc-enabled massive {MIMO}: A
  compressive sensing approach,'' \emph{{IEEE} Trans. Commun.}, vol.~66,
  no.~12, pp. 6164--6175, 2018.

\bibitem{Liang_TSP_2018}
L.~Liu and W.~Yu, ``Massive connectivity with massive {MIMO}—part i: Device
  activity detection and channel estimation,'' \emph{{IEEE} Trans. Signal
  Process.}, vol.~66, no.~11, pp. 2933--2946, 2018.

\bibitem{LiYang_TWC_2019}
Y.~Li, M.~Xia, and Y.-C. Wu, ``Activity detection for massive connectivity
  under frequency offsets via first-order algorithms,'' \emph{{IEEE} Trans.
  Wireless Commun.}, vol.~18, no.~3, pp. 1988--2002, 2019.

\bibitem{Unnikrishnan_TCOM_2021}
U.~K. Ganesan, E.~Björnson, and E.~G. Larsson, ``Clustering-based activity
  detection algorithms for grant-free random access in cell-free massive
  {MIMO},'' \emph{{IEEE} Trans. Commun.}, vol.~69, no.~11, pp. 7520--7530,
  2021.

\bibitem{Zhilin_TSP_2021}
Z.~Chen, F.~Sohrabi, and W.~Yu, ``Sparse activity detection in multi-cell
  massive {MIMO} exploiting channel large-scale fading,'' \emph{{IEEE} Trans.
  Signal Process.}, vol.~69, pp. 3768--3781, 2021.

\bibitem{WeiChen_TWC_2022}
W.~Chen, H.~Xiao, L.~Sun, and B.~Ai, ``Joint activity detection and channel
  estimation in massive {MIMO} systems with angular domain enhancement,''
  \emph{{IEEE} Trans. Wireless Commun.}, vol.~21, no.~5, pp. 2999--3011, 2022.

\bibitem{LiYang_TWC_2023}
Y.~Li, Q.~Lin, Y.-F. Liu, B.~Ai, and Y.-C. Wu, ``Asynchronous activity
  detection for cell-free massive {MIMO}: From centralized to distributed
  algorithms,'' \emph{{IEEE} Trans. Wireless Commun.}, vol.~22, no.~4, pp.
  2477--2492, 2023.

\bibitem{HaoZhang_TSP_2024}
H.~Zhang, Q.~Lin, Y.~Li, L.~Cheng, and Y.-C. Wu, ``Activity detection for
  massive connectivity in cell-free networks with unknown large-scale fading,
  channel statistics, noise variance, and activity probability: A bayesian
  approach,'' \emph{{IEEE} Trans. Signal Process.}, vol.~72, pp. 942--957,
  2024.

\bibitem{Tipping_2001_JMLR}
M.~E. Tipping, ``Sparse {B}ayesian learning and the relevance vector machine,''
  \emph{Journal of machine learning research}, vol.~1, no. Jun, pp. 211--244,
  2001.

\bibitem{Wipf_2004_TSP}
D.~Wipf and B.~Rao, ``Sparse {B}ayesian learning for basis selection,''
  \emph{{IEEE} Trans. Signal Process.}, vol.~52, no.~8, pp. 2153--2164, 2004.

\bibitem{Bishop_PRML}
C.~M. Bishop, \emph{{Pattern Recognition and Machine Learning}}.\hskip 1em plus
  0.5em minus 0.4em\relax {Springer New York}, 2006.

\bibitem{Alshoukairi_TSP_2018}
M.~Al-Shoukairi, P.~Schniter, and B.~D. Rao, ``A {GAMP}-based low complexity
  sparse {B}ayesian learning algorithm,'' \emph{{IEEE} Trans. Signal Process.},
  vol.~66, no.~2, pp. 294--308, 2018.

\bibitem{Marzetta_Fundamentals_2016}
T.~L. Marzetta, E.~G. Larsson, and H.~Yang, \emph{Fundamentals of massive
  {MIMO}}.\hskip 1em plus 0.5em minus 0.4em\relax Cambridge University Press,
  2016.

\bibitem{Ozlem_Cellfree_2021}
{\"O}.~T. Demir, E.~Bj{\"o}rnson, L.~Sanguinetti \emph{et~al.}, ``Foundations
  of user-centric cell-free massive mimo,'' \emph{Foundations and
  Trends{\textregistered} in Signal Processing}, vol.~14, no. 3-4, pp.
  162--472, 2021.

\bibitem{LTE2010b}
\emph{Further advancements for {E-UTRA} physical layer aspects (Release
  9)}.\hskip 1em plus 0.5em minus 0.4em\relax {3GPP} {TS} 36.814, Mar. 2010.

\bibitem{Jianan_WCOML_2022}
J.~Bai and E.~G. Larsson, ``Activity detection in distributed {MIMO}:
  Distributed {AMP} via likelihood ratio fusion,'' \emph{{IEEE} Wireless Comm.
  Letters}, vol.~11, no.~10, pp. 2200--2204, 2022.

\end{thebibliography}
	\end{document}